\documentstyle[12pt,epsfig]{article}
\def\beq{\begin{equation}}   \def\eeq{\end{equation}}
\def\bea{\begin{eqnarray}}   \def\eea{\end{eqnarray}}

\newcommand{\gsim}{\lower.7ex\hbox{$
\;\stackrel{\textstyle>}{\sim}\;$}}
\newcommand{\lsim}{\lower.7ex\hbox{$
\;\stackrel{\textstyle<}{\sim}\;$}}

\newcommand{\Lam}{\Lambda_{\rm QCD}}

\renewcommand{\Im}{\mbox {Im}\:}

\newcommand{\bibit}[1]{\bibitem{#1}}

\begin{document}

\def\lsim{\mathrel{\rlap{\lower3pt\hbox{\hskip0pt$\sim$}}
    \raise1pt\hbox{$<$}}}         
\def\gsim{\mathrel{\rlap{\lower4pt\hbox{\hskip1pt$\sim$}}
    \raise1pt\hbox{$>$}}}         

\begin{titlepage}
\renewcommand{\thefootnote}{\fnsymbol{footnote}}

\begin{flushright}
JLAB-THY-99-06\\
UND-HEP-99-BIG\hspace*{.2em}03\\
hep-ph/9903258\\
\end{flushright}
\vspace{1.1cm}

\begin{center} \Large
{\bf
Pauli Interference in the 't Hooft Model:\\ 
\hspace*{-.8em} $\mbox{\bf Heavy Quark Expansion and Quark-Hadron Duality}$ 
}
\end{center}
\vspace*{.8cm}
\begin{center}
{\Large
Ikaros Bigi$^{\:a}$ and Nikolai Uraltsev$^{\:a-c}$
\\
\vspace{.5cm}
{\normalsize
$^a${\it Dept.\ of Physics,
Univ.\ of Notre Dame du
Lac, Notre Dame, IN 46556, U.S.A.}\\
$^b${\it Jefferson Lab, 12000 Jefferson Avenue, Newport News, VA
23606}\\
$^c${\it Petersburg Nuclear Physics Institute,
Gatchina, St.\,Petersburg, 188350, Russia
}
}
}

\vspace*{3.2cm}
{\large \bf Abstract}
\vspace*{.25cm}

\end{center}

\thispagestyle{empty}
\setcounter{page}{0}

\noindent
Pauli Interference in decays of heavy flavor mesons, a genuinely
nonleptonic nonperturbative effect, is considered in the 't~Hooft 
model. Analytic summation of the exclusive decay widths yields the same
expression as obtained in the OPE-based approach. Threshold-related
violations of local duality in PI are found to be suppressed. A novel
case is identified where the OPE effect is completely saturated by a
single final state.

\end{titlepage}
\setcounter{footnote}{1}

\newpage

The dynamics driving heavy flavor transitions are concisely 
expressed on the quark level in terms of CKM parameters and masses. Yet
their impact on the observable  weak decays of heavy flavor hadrons 
$H_Q$ is obscured by nonperturbative features of strong interactions. 
The principal elements of the model-independent treatment of 
inclusive decay probabilities based on Wilson's operator product
expansion (OPE) \cite{wilson}, were laid down almost 15 years ago
\cite{vsold,volshif}. 
Significant progress has later been achieved in developing this
technique and establishing theoretical
control over nonleptonic, semileptonic and radiative channels of
practical interest (for a recent review, see Refs.\,\cite{reviews}).
The nonperturbative corrections which would scale like $1/m_Q$ are absent
from the truly inclusive widths; the leading nonperturbative
corrections arise at order $1/m_Q^2$ and are ``flavor-independent'' 
corrections  insensitive to the flavor of the spectator(s) 
\cite{buv,bs}.
Effects sensitive to the flavor of the spectator(s) emerge 
at the $1/m_Q^3$ level \cite{vsold,miragewa,buv}.  
They are conventionally called Weak Annihilation
(WA) in mesons \cite{oldwa}, Weak Scattering (WS) \cite{tramp} in baryons and 
Pauli Interference (PI) \cite{guber} in both systems. OPE relates their 
contributions to the expectation values of local four-quark operators.

The OPE is typically constructed as an asymptotic power expansion in 
Euclidean space, and neglects the terms $\sim \! \exp{(\!- \sqrt{Q^2}/\Lam)}$ 
suppressed for high momentum scales $\sqrt{Q^2}$. Yet in
Minkowski space these can lead to terms which oscillate and have
only power suppression. For heavy quark decays the practical OPE 
{\it a priori} does
not forbid, for instance, dependence of the type
$\sin{(m_Q\Lam)}/m_Q^k$ \cite{shifcont,inst}. 
To which degree they are suppressed by powers of $1/m_Q$ depends on details 
of the strong forces and the specifics of the process under study. 
This problem is behind the violation of local quark-hadron duality. 
Since WA and PI represent power suppressed effects, one 
might expect numerically larger corrections to duality at 
finite values of $m_Q$ than for the leading terms in the OPE. 
Even a more radical concern had been expressed in the past that the OPE 
is not applicable to nonleptonic heavy flavor decays at all, even at arbitrary
large heavy quark mass.

Implementation of duality in nonleptonic decays is not always obvious at 
first glance, in particular for PI. The underlying quark diagram for 
PI in $B^-$ mesons
is shown in Fig.\,1. The $\bar u$ antiquark produced in the decay must be 
slow to  interfere with the valence $\bar u$. Tracing the color flow 
one is then faced with a dichotomy. 
The large momentum flows through the diquark loop ($cd$) which 
represents the ``hard core'' of the process. In practical OPE 
one effectively replaces the propagation of this diquark by a 
nearly free di-fermion loop; the absorptive part is evaluated -- at
least to leading order -- via the production of free quarks.
Such identification cannot be justified by itself even at arbitrarily
large momentum transfer, since the spectrum of QCD does not contain colored 
states.

\thispagestyle{plain}
\begin{figure}[hhh]
 \begin{center}
 \mbox{\epsfig{file=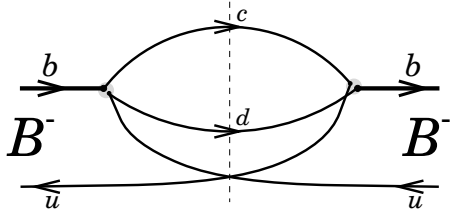,width=6cm}}
 \end{center}
 \caption{
Quark diagram for PI in $B^-$ decays. The weak vertices are broken
to show the color flow for the leading-$N_c$ contribution.
}
\end{figure}

Alternatively, one can combine a hard quark $q$ from the loop with 
the slow valence antiquark to arrive at a color singlet 
configuration possibly 
dual to the hadronic spectrum. That 
comes at a price, though: this $q \bar q_{sp}$ system is 
no longer a hard one. For at least in a partonic picture with 
$p_{\rm sp} \sim m_{sp}\! \to \! 0$ the $q \bar q_{sp}$ invariant 
mass vanishes irrespective of $m_b$.

An intriguing example that interference might be delicate was 
noted in Ref.\,\cite{shifman}, where an inclusive and exclusive
treatment of $B$ decays seemed to yield different results for
the interference of two separate color operators.

A dedicated consideration suggests that such apparent
puzzles can be resolved without invoking new paradigms or 
assuming a higher onset of duality, growing with $N_c$.
Here we will present an explicit analysis of PI 
in a soluble field theory, the `t~Hooft model \cite{thooft} which is 
QCD in 1$+$1 dimensions 
in the limit $N_c \!\to\! \infty$. We have found 
that the OPE expression matches the result obtained 
when summing over the hadronic resonances; 
the onset of duality is largely independent of the details. 
Moreover, the parton-deduced OPE
expression appears to be {\em exact} in the chiral limit when all
involved quarks except $Q$ are massless. In this case the considered
effect is saturated by a single final state. From this perspective it is
complementary to the classical small velocity limit in semileptonic
decays \cite{volshif} where a similar saturation holds in the actual 
$D\!=\!4$ QCD for heavy enough $b$ and $c$ quarks, up to power corrections.

The 't Hooft model has already been employed 
recently for studying inclusive 
heavy flavor decays, both analytically \cite{D2} 
and numerically \cite{gl12}. The large-$N_c$ limit 
harnessed in the analyses, however, washed out the difference between 
semileptonic
and nonleptonic decays. In the case of weak annihilation, additionally,
it equates the validity of the OPE for the heavy flavor width with its
applicability to the vacuum current-current correlators, and thus can 
be thought not to be representative for the problems encountered in the 
nonleptonic decays. PI 
is a genuinely nonleptonic effect. Invoking the spectator quark 
makes this a leading-$N_c$ effect and allows
a consistent evaluation in the framework of the 't~Hooft eigenstate
problem. Presentation of full analysis and details of the heavy quark
expansion in the model is left for a separate 
dedicated publication.
\vspace*{.35cm}

{\large \bf The `t Hooft model and heavy quark decays}

\vspace*{.2cm}
\noindent
The `t~Hooft model -- QCD in 1$+$1 dimensions 
with $N_c \!\to \!\infty$ -- has been described in
many papers \cite{thooft,ff}. While the Lagrangian has the usual form
\begin{equation}
{\cal L}_{1+1}=-\frac{1}{4g_s^2} \,
G_{\mu\nu}^a G_{\mu\nu}^a \,+\, \sum
\bar\psi_i
(i\not \!\!D -m_i)\psi_i \; , \; \;\;\;
i D_\mu=i\partial_\mu + A_\mu^a T^a\, ,
\label{9}
\end{equation}
the coupling $g_s$ has dimension of mass making the theory 
superrenormalizable. $A_\mu$ has dimension of mass as in $D=4$ in the
adopted normalization; however the fermion fields $\psi(x)$ carry 
dimension of $m^{1/2}$. In the large $N_c$ limit strong interaction
effects are driven by the parameter 
\begin{equation}
\beta^2\;=\;\frac{g_s^2}{2\pi}\,\left( N_c -\frac{1}{N_c} \right)
\label{beta}
\end{equation}
playing the role of the nonperturbative scale $\Lam$. Additional
benefit of the $D=2$ theory is that the observables can be expressed in 
terms of {\em bare} masses and couplings, and that the perturbative
corrections for hard quantities generate power series. More relevant
details and discussions can be found in Refs.\,\cite{D2,d2wa}.

In the limit $N_c \!\to \! \infty$ the spectrum of 1$+$1 QCD consists of
mesonic quark-antiquark bound states which are stable under strong
interactions. The meson masses are given by eigenvalues of the 't~Hooft
equation
\begin{equation}
M^2_{n}\varphi_n(x) = \left[
\frac{m_1^2 - \beta ^2}{x} + \frac{m_2^2 - \beta ^2}{1-x}
\right]\varphi_n(x) - \beta ^2 \int_0^1{\rm d}y 
\,\frac{\varphi_n(y)}{(y - x)^2}\;,
\label{30}
\end{equation}
where $m_{1,2}$ are the masses of the quark constituents, and the
integral is understood in the principal value prescription.
The solutions are the light-cone wave functions $\varphi_n(x)$.

Polarization operator of the vector $\bar{d}\gamma_\mu u$ current 
\begin{equation}
\Pi_{\mu\nu}(q^2) \;=\;
\frac{1}{\pi}\: \Pi(q^2)\,
\left(q^2\delta_{\mu\nu}-q_\mu q_\nu \right)\;,
\qquad
\rho(q^2)\;\equiv \;
-\frac{1}{\pi}\:\Im \Pi(q^2)
\label{18}
\end{equation}
(assuming that $m_u\!=\!m_d$ 
and the current is strictly
conserved) has a resonance form
\begin{equation}
\Pi(q^2)\;=\; \pi \sum_n \,\frac{f_n^2}{q^2-M_n^2}\;, \qquad
\rho(q^2)\;=\; \pi \sum_n \,f_n^2\,\delta\left(q^2-M_n^2\right)\;,
\label{35}
\end{equation}
where the decay constant $f_n$ of a particular meson $n$ is given by
\begin{equation}
f_n\;=\; \sqrt{\frac{N_c}{\pi}}\,
\int_0^1 {\rm d}x \:\varphi_n(x)\;.
\label{34}
\end{equation} 
In $D=2$ the exact correlator of vector currents for {\em massless} quarks
is very simple:
\begin{equation}
\Pi(q^2) \;=\;
\frac{N_c}{q^2}
\;,
\qquad
\rho(q^2)\; = \; N_c \,\delta(q^2)\;.
\label{22}
\end{equation}
In the 't~Hooft model at $m_{u,d}\!=\!0$ one has 
$f_0\!=\!\sqrt{N_c/\pi}$ and
$M_0\!=\!0$; for all excitations $f_n\!=\!0$.
With nonzero quark masses the spectral density shifts upward, to the
mass scale $\sim \!\beta m_{u,d}$ or $m_{u,d}^2$, 
and a high-energy tail in $\rho$
appears $\sim \!N_c(m_u^2+m_d^2)/q^4$ \cite{d2wa}.

We choose the weak decay interaction of 
the current-current form. Since in 1$+$1 dimensions 
the axial current is related to the vector one, 
$J_\mu^A \!=\! \epsilon_{\mu\nu} J_\nu^V$, we simply consider 
the $V$$\times$$V$ interaction, and include both charge- and
neutral-current vertices:
\begin{equation}
{\cal L}_{\rm weak} \;=\;-\frac{G}{\sqrt{2}}\,
\left(\, a_1(\bar c \gamma_\mu b)\, (\bar{d}\gamma^\mu u )\; + \;
a_2(\bar d \gamma_\mu b)\,(\bar{c}\gamma^\mu u ) \, \right)
\; + \; {\rm H.c.}\;.
\label{95}
\end{equation}
While $G$ is an analogue of the Fermi constant, it is dimensionless 
here. 
In the example of the parton width 
the interference term $\propto \! 2a_1 a_2$ is non-leading in $1/N_c$: 
\begin{equation}
\Gamma^{\rm quark}\;=\;-\frac{G^2}{2} N_c\,
\left(\, a_1^2 \Gamma_1^{\rm quark} +
a_2^2 \Gamma_2^{\rm quark} + 
2a_1 a_2 \Gamma_{12}^{\rm quark}\right)\;,
\label{96}
\end{equation}
where $\Gamma_{1,2}^{\rm quark}$ are $\sim \! {\cal O}(N_c^0)$, 
while $\Gamma_{12}^{\rm quark} \!\sim \!
{\cal O}(1/N_c)$. Yet also the interference term can be leading 
in $1/N_c$ \cite{ruckl} -- if the contributions are non-leading in $1/m_b$. 
Specifically, considering $B^-$ decays 
affected by PI, see Figs.\,1 and 2, we find  
$\Gamma_{12}^{\rm PI} \!\sim \! {\cal O} (N_c^0) 
$;
the three terms $\Gamma_1$, $\Gamma_2$ and 
$\Gamma_{12}^{\rm PI}$ are all of the 
same order in $1/N_c$. This allows one to retain only the leading-$N_c$ 
decay amplitudes.

\thispagestyle{plain}
\begin{figure}[hhh]
 \begin{center}
 \mbox{\epsfig{file=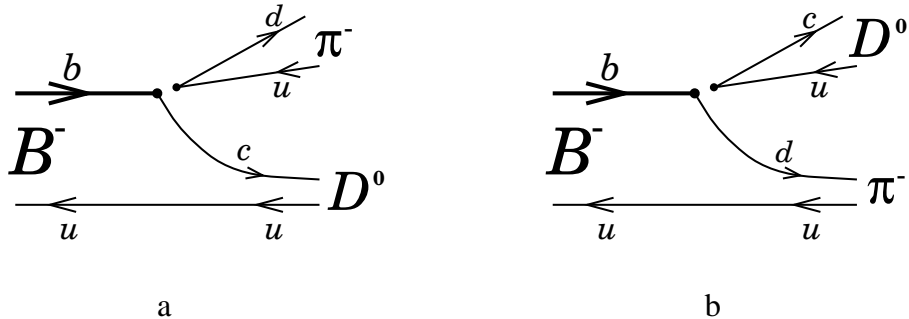,width=12cm}}
 \end{center}
 \caption{
Large-$N_c$ decay amplitudes induced by charge-current (a) and
neutral-current (b) terms in the weak decay Lagrangian.
}
\end{figure}

There is something else we can read off from Fig.\,2. 
The two quark-antiquark clusters in the final state are 
produced with quite different characteristics: 
while the upper one is produced by a ``pointlike" 
source, the lower is not, since it combines a decay 
quark with the valence quark; we will refer to 
the latter as 
``multiperipheral" production. On general grounds one 
expects the typical mass distribution for hadrons from the upper 
vertex to be considerably different from that from the lower one.
Normally, the former is much wider spanning over the whole accessible
range up to $m_b$. However, for the special case of vector-like
interactions in $D\!=\!2$ the bulk of pointlike-produced hadrons has 
smaller masses comparable to the corresponding quark masses.
Yet interference between the two diagrams in Fig.\,2 requires the 
$d \bar u$ and $c \bar u$ clusters to have the same mass in 
both cases. This explains 
why PI is power suppressed -- a property that emerges 
automatically in the OPE. 

We turn now to the 't Hooft model to transform these qualitative 
observations into specific expressions.  
Fig.\,1 generates -- to leading order in the OPE -- the operator 
$$
\hat{\Gamma}^{\rm PI} = \!
- a_1 a_2
\frac{G^2}{K}\left\{ \!
\left(1\!-\!\frac{m_c^2\!+\!m_d^2}{m_b^2}\right)\!
(\bar b \gamma_\mu \!\gamma_5 u) (\bar u \gamma_\mu \!\gamma_5 b)\!
-\!
\frac{2m_c m_d}{m_b^2}\!
\left[(\bar b u) (\bar u b)\!+\!
(\bar b i\gamma_5 u) (\bar u i\gamma_5 b) \right]\!
\right\} \!,
$$
\begin{equation}
K= \left[ \left( 1 - \frac{(m_c + m_d )^2}{m_b^2} \right) 
\left( 1 - \frac{(m_c - m_d )^2}{m_b^2} \right) \right] ^{1/2}\;.
\label{99}
\end{equation}
This contribution remains finite for 
$m_c \!=\! m_d \!=\! m_u\! =\!0$ and can conveniently be analyzed in 
this chiral limit. 
The $B^-$ expectation value of the operators in Eq.\,(\ref{99})
has a $N_c$-favorable color structure and is, therefore
given by vacuum factorization: 
\beq
\label{100}
\frac{1}{2M_B} \langle B^-|
(\bar b \gamma_\mu \gamma_5 u)\,(\bar u \gamma_\mu \gamma_5 b)
|B^- \rangle
\; = \;  \frac{1}{2}\, f_B^2 M_B
\;,
\eeq
thus yielding in the chiral limit 
\begin{equation}
\Gamma^{\rm PI}\;= \; -2 a_1 a_2\,  \frac{G^2}{4}\; f_B^2 M_B\;.
\label{101}
\end{equation}
We note that $\Gamma^{\rm PI}$ approaches a constant for 
$m_b \!\to \! \infty$ since $f_B \!\propto \!1/\sqrt{m_b}$. It constitutes a
$1/m_Q$ nonperturbative correction to the decay width, in contrast to
$1/m_Q^3$ in four dimension. This follows from different
canonical dimension of quark fields \cite{D2}.

There are no $1/m_b$ corrections to this result.
This is a peculiarity of two dimensions where the absorptive part of the
(di)quark loop in  Fig.\,1 scales as the momentum to the zeroth power 
and, thus does not depend on
whether quark or meson momentum 
flows through it. The
corrections to the Wilson coefficient or other higher-order operators
can induce only terms suppressed by at least two powers of 
$1/m_b$, if $m_{c,u,d} \ll m_b$.

Let us now consider the decays in terms of hadrons. 
To leading order in $1/N_c$ the final states are pairs of mesons
generically referred to as $D^0_k$ and $\pi^-_n$. 
The partial decay width $B\to D^0_k \pi^-_n$ takes the form
\begin{equation}
\Gamma_{kn}\;= \; \frac{G^2}{8 M_B^2 |\vec{p}\,|}\:
\left[\,
a_1^2|{\cal A}_k {\cal B}_n|^2\, +\, a_2^2|{\cal A}_n {\cal B}_k|^2 \,+ \,
2 a_1 a_2 \,{\rm Re}\, {\cal A}_k {\cal A}_n^* {\cal B}_k^* {\cal B}_n \,
\right]\;,
\label{102}
\end{equation}
where ${\cal A}$ and ${\cal B}$ schematically denote the ``multiperipheral''  
transition amplitudes and the ``pointlike'' meson creation amplitudes,
respectively:
\begin{equation}
{\cal A}_k \;\sim \; \langle k|J_\mu |B \rangle\;, \qquad \;
{\cal B}_n \;\sim \; \langle n|J_\mu |0 \rangle\;.
\label{103}
\end{equation}
With $\vec{p}$ denoting the rest-frame momentum of the final state
mesons, the PI width is then given by the sum 
\begin{equation}
\Gamma^{\rm PI}\;= \; 2 a_1 a_2\,\frac{G^2}{8 M_B^2}\:
\sum_{k,n}\,\frac{1}{|\vec{p}\,|} \,{\cal A}_k {\cal A}_n^* {\cal B}_k^* 
{\cal B}_n
\;.
\label{104}
\end{equation}
Eqs.\,(\ref{102},\ref{104})
reflect the peculiar 
two-body phase space $\propto \! 1/|\vec{p}\,|$ in two dimensions.

The quantitative matching between the OPE-based calculation and the
hadronic saturation of the interference width is most easily seen when
all final state quarks $u,d,c$ are massless. While not affecting the OPE
analysis, this limit significantly simplifies the expressions for the
individual hadronic amplitudes, as explained in Refs.\,\cite{D2,d2wa}. In the
case at hand only $n=0$ survives for the decay amplitude $\sim a_1$
(Fig.\,2a) and $k=0$ for the amplitude $\sim a_2$ (Fig.\,2b). The
interference then resides in the single final state containing the
lowest lying massless $D^0$ and $\pi^-$.  Moreover, the transition
amplitudes between the two  mesons take a  particularly simple form at
$q^2=0$ \cite{D2} in terms  of their `t~Hooft wavefunctions: 
\begin{equation}
q_\mu\:\frac{1}{2M_l} \langle k|\epsilon_{\mu\nu} J_\nu |H_Q \rangle
\;=\;
-q_z\, \int_0^1 \, {\rm d}x \: \varphi_k(x) \varphi_B(x)
\;,
\label{105}
\end{equation}
with the `pointlike' amplitude given by 
${\cal B}_0\sim i\epsilon_{\mu\nu} f_0 P_\nu^{(0)}$ yielding 
\begin{equation}
\Gamma^{\rm PI}\;= \; -2 a_1 a_2\,\frac{G^2 M_B N_c}{4\pi}\:
\left|\int_0^1 \, {\rm d}x \: \varphi_B(x)\right|^2\;= \;-
2 a_1 a_2\,\frac{G^2}{4} \, f_B^2 M_B\;,
\label{106}
\end{equation}
where we have used the fact that $\phi_0(x)\!=\!1$
for massless
quarks. The minus sign emerges since the direction of the 
vector playing the role of $\vec q$ is opposite for the two 
interfering amplitudes. 

Thus, the OPE prediction Eq.\,(\ref{95}) is exactly reproduced.
Apparently, there is no violation of local duality at all for PI
when $m_u\!=\!m_d\!=\!m_c\!=\!0\,$! It could be anticipated, for 
in this limit the
only threshold in $\Gamma^{\rm PI}$ occurs at $m_b\!=\!0$.
Then the OPE series can have the same convergence properties in Minkowskian 
as in Euclidean space \cite{shifcont}.

With $m_{u,d,c}\!\ne\! 0$ the situation becomes more complex since 
more production thresholds arise. Those exhibit singularities 
due to the singular two-body phase space 
in $D\!=\!2$. As explained in detail elsewhere, the predictions
of the practical OPE must be compared to the decay probabilities where
the threshold singularities are averaged, and for heavy quark decays
this procedure is naturally done by smearing over the interval of the
heavy quark mass, exceeding the distance between the successive
principal thresholds \cite{shifcont,d2wa}.
Since this phase space factor is integrable, the threshold spikes do not 
affect the width smeared over a mass interval $\Delta m_b \sim
1/m_b$. 

The calculation uses the detailed kinematic duality
between the partonic and hadronic probabilities. Details of the
derivation for the massive case can be found in Ref.\,\cite{d2wa}. Here
we only sketch the basic steps. First, one takes into account that for
the bulk of the decays the 
masses of both `pointlike' $i$ and `multiperipheral' $j$ mesons are small
compared to $m_b$:
\beq
M_i^2\, \lsim \,m_q^2,\, \beta m_q\,;\qquad 
M_j^2 \, \lsim \, \beta m_b\,,
\label{pi10}
\eeq
where $m_q$ generically denotes the final state quark masses. These
features anticipated from the quark-level consideration has been
demonstrated directly in the 't~Hooft model in Refs.\,\cite{D2,d2wa},
where their accuracy was also quantified. It is sufficient for
calculating $\Gamma^{\rm PI}$ up to terms $1/m_b^2$.
Then we can expand momentum $|\vec p\,|$ and transition 
amplitudes around the massless kinematics:
\begin{equation}
\frac{1}{|\vec{p}\,|} = \frac{2}{M_B} \left(
1+2\frac{M_k^2+M_n^2}{M_B}\,+\,...
\right) \; , \; \; \;
{\cal A}_k(M_n^2) \simeq {\cal A}_k(0) +\frac{M_n^2}{m_b^2}\, 
m_b^2 \left.\frac{{\rm d}A_k}{{\rm d}q^2}\right|_{q^2=0}\;.
\label{108}
\end{equation}
The set of sum rules derived in Ref.\,\cite{D2} allows to calculate
the sum of the decay probabilities and the leading corrections
associated with nonzero meson masses; the latter cancel the increase in
$1/|\vec p\,|$ in Eq.\,(\ref{104}).

For example, neglecting deviation of $q^2$ from $0$ in calculating the
`multiperipheral' amplitudes, we get for the two interfering amplitudes
\cite{D2,d2wa}
\bea
\nonumber
& {\cal A}_k {\cal B}_n \;=\; &\sqrt{\frac{N_c}{2\pi}}\, (M_B^2\!-\!M_k^2)\:
\int_0^1 \, {\rm d}x \: \varphi_B(x) \varphi_k(x)\cdot
\int_0^1 \, {\rm d}y \: \varphi_n(y)\\
& {\cal A}_n {\cal B}_k \;= -& \sqrt{\frac{N_c}{2\pi}}\, (M_B^2\!-\!M_n^2)\:
\int_0^1 \, {\rm d}x \: \varphi_B(x) \varphi_n(x)\cdot
\int_0^1 \, {\rm d}y \: \varphi_k(y)\;.
\label{110}
\eea
Extending summation over all $k$ and $n$ and using the completeness of
the 't~Hooft eigenstates, we obtain
\begin{equation}
\Gamma^{\rm PI}\;= \;\sum_{k,n}  \Gamma^{\rm PI}_{kn} \,\simeq\,
-2a_1a_2\, \frac{G^2 N_c}{4\pi M_B^2}
\left| \int_0^1 \, {\rm d}x \: \varphi_B(x) \right|^2\,=\,
-2 a_1 a_2\, \frac{G^2}{4}\: f_B^2
\;.
\label{111}
\end{equation}
This expression is valid up to ${\cal O}(1/m_b^2)$ corrections. The
leftover effect of the slope of the transition formfactor is quadratic
in $m_q/m_b$ \cite{d2wa}.

Thus, $\Gamma^{\rm PI}$ {\em agrees with the expression given by
the free quark loop} to the accuracy suggested by the OPE.

For small, yet nonvanishing final state quark masses one can 
estimate the limitation in local duality for PI as it relates to 
thresholds. Close to the opening of such a threshold the hadronic 
width is not exactly reproduced by the OPE. These effects 
are characterized by the nature of the threshold singularity, the scaling of 
its residue with $m_b$ and the distance between principal thresholds. 

It turns out that these effects essentially depend on the 
relation between the final state masses. The strongest 
violation of local duality occurs when one 
of the mesons is a low excitation while the other has a 
large mass close to $m_b$. 

The case with $m_c \!\ne\! 0$, $m_u \!= \!m_d\! =\! 0$ is somewhat special. 
For one of the interfering decay amplitudes vanishes right 
at threshold and $\Gamma ^{\rm PI}$ undergoes a finite 
jump there (details can be found in Ref.\,\cite{d2wa}): 
\begin{equation}
|\delta \Gamma^{\rm PI}_{k0}|
\propto
\frac{G^2}{2}\, 2a_1a_2  \, f_B \,m_c
\frac{\beta^{9/2}}{M_B^{9/2}}\: \theta (M_B\!-\!M_k)
\;, \qquad 
M_{k+1} \!- \!M_k \simeq \frac{\pi ^2 \beta ^2}{2M_B}\;.
\label{115}
\end{equation}
A more elaborate analysis suggests that the relative sign 
of the 
two interfering amplitudes alternates for successive thresholds, and 
we arrive at the following ansatz: 
\begin{equation}
\delta \Gamma^{\rm PI}_{\rm osc}
\;\propto \;
\frac{G^2}{2}\, 2a_1a_2 \, \,f_B \,m_c
\frac{\beta^{9/2}}{M_B^{9/2}}\: \sum_k (\!-\!1)^k \,
\theta \left(M_B\!-\!\pi\beta\sqrt{k}\right)
\;.
\label{117}
\end{equation}
The oscillating term in PI thus gets damped by at least $1/m_b^5$. 

For $m_{u,d}$ nonzero, the picture changes essentially in two 
respects: neither decay amplitude vanishes at threshold, however 
the phase space factor softens to $[2m_\pi(M_B - M^{(k)}_{\rm thr})]^{-1/2}$
for the decays $B\to D^{(k)}+\pi$ at $m_\pi\ne 0$.
One then finds 
\beq 
\delta \Gamma^{\rm PI}_{\rm osc}
\propto 
\frac{G^2}{2}\, 2a_1a_2\,
\frac{ m_c m_\pi^{1/2}\beta^{5}}{M_B^{5}}\: \sum_k (\!-\!1)^k \,
\frac{\theta \left(M_B\!-\!M_{\rm thr}^{(k)}\right)}
{\sqrt{M_B\!-\!M_{\rm thr}^{(k)}}}
\:, \qquad
M_{\rm thr}^{(k)} \simeq  m_\pi \!+\! \pi \beta\sqrt{k} 
\:.
\label{118}
\end{equation}
With all final-state masses non-zero, many additional 
thresholds open up and the dynamical landscape becomes in general 
rather complex. Yet one observes that the contributions from the final
states where both meson masses constitute a finite fraction of $m_b$ 
are suppressed by even higher powers of $1/m_b$ \cite{d2wa}. 
\vspace*{.35cm}

{\large \bf Conclusions}
\vspace*{.2cm}

\noindent
Let us first describe the hadronization picture 
suggested by the large-$m_b$ analysis. We observe that duality applies
to the quark-antiquark pairs where each energetic quark ($c$ or
$d$) is combined with the wee spectator antiquark or slow
$\bar{u}$ produced in the weak vertex. The completeness of the hadronic
states -- or, in other words, the duality between the parton-level and
mesonic states -- is achieved already for a single fast moving decay
quark when it picks up a slow spectator. The `hardness' of
these processes determining the applicability of the quasifree
approximation, depends on the energy of the fast quark rather than
on the invariant mass of the pair.
For actual hadrons, the slow spectator(s) compensate the
color of the initial static heavy quark. This initial distribution of
the color field singles out the rest frame and introduces the new scale,
energy {\it vs.}\ invariant mass.

In the full graph for the meson decay which would explicitly include
propagation of the spectator, there is always a color mate
for any quark in the ``partonic'' part of the diagram. Therefore,
considering the colored diquark loop as nearly free yields the correct
result for the properly formulated problem.

In this note we have analyzed PI as a spectator-dependent nonleptonic
effect 
for heavy flavor decays within the 
't~Hooft model. We have found the OPE expression to match 
the result obtained when summing over exclusive final states 
at least up to terms $(\Lam/m_b)^3\,\Gamma_B$ which is the  
accuracy we adopted. 
This is the first analysis where the
peculiarities of nonleptonic decays widths of heavy flavors emerge 
compared to their semileptonic decays or processes 
like ${\rm e}^+{\rm e}^-$ annihilation. 
We did not encounter specific problems 
for applying the OPE, in accord with the general
arguments justifying it for all types of fully inclusive decay widths
\cite{miragewa,inst,D2,d2wa}.

We estimated the effects of violation of local duality in PI in the heavy
quark limit. It turned out to be strongly suppressed scaling like
$1/m_b^5$. A novel case was identified where local parton-hadron duality
is exact. Although applicable only in two dimension for large $N_c$, it
complements the classical case of small velocity semileptonic $b\to c$
decays \cite{volshif} which was instrumental in developing the modern
applications of the heavy quark expansion.
\vspace*{.2cm}

{\bf Acknowledgments:} \hspace{.1em} 
We are indebted to 
M.\ Shifman and A.\ Vainshtein for illuminating insights and their
encouraging interest.
This work was supported in part by the National Science Foundation under
the grant number PHY~96-05080 and by the RFFI under the grant number
96-15-96764.

\end{document}